\input{aipcheck}

\documentclass[
    ,final            
  ]
  {aipproc}

\layoutstyle{6x9}

\usepackage{amsmath}
\usepackage{wrapfig}
\usepackage{amssymb}
\usepackage{lineno}
\usepackage{color}
\usepackage{graphicx}
\usepackage{relsize}
\def\babar {{{\mbox{\slshape B\kern-0.1em{\smaller A}\kern-0.1em B\kern-0.1em{\smaller A\kern-0.2em R }}}}}

\begin{document}

\title{Results on Charmonium-like States at \babar}


\author{Valentina Santoro }{
  address={INFN Ferrara, via Saragat 1, 44122 Ferrara, Italy }
}

\begin{abstract}
 We present recent results on charmonium and charmonium-like states from the BaBar B-factory located at the PEP-II asymmetric energy $e^{+}e^{-}$ storage ring at the SLAC National Accelerator Laboratory. \end{abstract}

\maketitle


\section{Introduction}

  The charmonium spectroscopy has been revitalized by the discovery of many new states above the open charm threshold. While some of them
  appear to be consistent with conventional charmonium others don't seem to behave like standard meson and could be made of a larger number of constituent quarks. While this possibility has been considered since the beginning of the standard model, the clear identification of such states would represent a major revolution in the picture of the quark model. The possible interpretations of these states beyond the conventional mesons are:
  \begin{itemize}
  \item molecules: these are loosely bound states of a pair of mesons $([Q\bar{q}][q^{'}\bar{Q}]$) \cite{mol}. This system would be stable if the binding energy 
  were to set the mass of the states below the sum of the two meson masses. The dominant binding mechanism of this state is pion exchange. 
  \item tetraquarks: a bound state of four quarks usually represented as  $([Q q][\overline{q^{'}Q}]$) \cite{mai}. Strong decays
proceed via rearrangement processes. This interpretation predicts many new states and the existence of states with non-zero charge.    
  \item hybrids: a bound state of a quark-antiquark pair and a number of constituent gluons. The lowest-lying state from lattice calculation is expected have mass about 4.1 $\mathrm{GeV/c^{2}}$\cite{hyb}. While states with exotic quantum number such $(0^{+-},~1^{-+},~2^{+-})$ are unambiguos hybrids other states can be distinguished from the conventional hadrons by characteristics of their decay process.
  \end{itemize}
  In addition there is the possibility that some observed enhancements are simply due to threshold effect: a given amplitude might be magnified when new hadronic final states become energetically available.\\
  Even if several experiments \cite{qr} have performed many measurements of these charmonium-like states the overall picture is not clear. In the next section we will review recent \babar results on this field.
 
\section{Charmonium production @ the B-factories}
The B-factories are an ideal place to study charmonium since charmonium states are copiously produced in a variety of processes: 
\begin{itemize}
\item B decays: B mesons decay to charmonium in about 3\% of cases. Charmonium states of any quantum number can be produced.
\item Double-charmonium production: in this process, observed for the first time by Belle \cite{Abe:2002rb}, a $J/\psi$ or a $\psi(2S)$ is
produced together and exclusively with another charmonium state.
\item  $\gamma \gamma$ fusion:  
two virtual photons are emitted by the colliding
$e^{+}e^{-}$ pair $(e^{+}e^{-}\to e^{+}e^{-} \gamma^{*} \gamma^{*}\to e^{+}e^{-} (c\bar{c})$). States with C=+1 are formed.
\item  Initial state radiation (ISR): where a photon
is emitted by the incoming electron or positron
$(e^{+}e^{-}\to \gamma (c\bar{c}))$, only  states with $J^{PC}=1^{--}$ are formed.
\end {itemize}

\section{Study of the $J/\psi \omega$ in two-photon interactions.}
The Y(3940) was observed for the first time by Belle \cite{Abe:2004zs} in B decays
 and then confirmed by \babar  \cite{babary3940)}. In a re-analysis  \cite{X3872} of the \babar data sample the precision of the Y(3940) parameters was improved and in addition evidence for the decay $X(3872) \to J/\psi \omega$ was found. This confirmed an earlier Belle claim  \cite{X3872belle}
 for the existence of this decay mode. A subsequent Belle paper \cite{X3915belle} reports the evidence of a structure in the process $\gamma \gamma \to J/\psi \omega$ that they dubbed the  X(3915) with mass and width values similar to those obtained for the Y(3940) by \babar \cite{babary3940)}. \babar has recently performed a study of the process  $\gamma \gamma \to J/\psi \omega$~\cite{X3915BaBar}
 to search for the X(3915) and the X(3872) using a data sample of 519 $\mathrm{fb^{-1}}$. Figure 1 shows the reconstructed $J/\psi \omega$ mass distribution after all the selection criteria have been applied. A large peak at near $3915~{\mathrm MeV/c^{2}}$ is observed with a significance of 7.6 $\sigma$. The measured resonance's parameters obtained from a maximum likelihood fit are $m_{X(3915)}=(3919.4 \pm 2.2  \pm 1.6) {\mathrm MeV/c^{2}}$, $\Gamma_{X(3915)}=(13\pm 6\pm 3)$MeV. The measured value of the two-photon width times the branching fraction, is $\Gamma_{\gamma \gamma}(X(3915) \times {\cal B}(X(3915) \to J/\psi \omega) =52\pm 10 \pm 3$~eV and $(10.5 \pm 1.9 \pm 0.6)$~eV for the spin hypothesis J=0 and J=2, respectively, where the first error is statistical and the second is systematic. In addition a Bayesian upper limit (UL) at 90 \% confidence level (CL) is obtained for the X(3872): $\Gamma_{\gamma \gamma}(X(3872) \times {\cal B}(X(3872) \to J/\psi \omega <1.7 $~eV, assuming J=2.

\begin{figure}
  \includegraphics[height=.22\textheight]{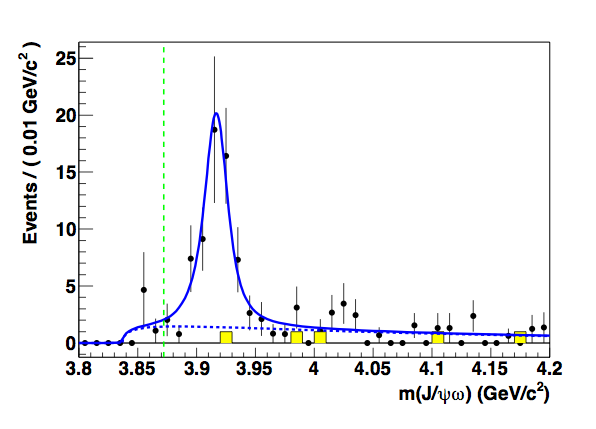}
  \label{fig1}
  \caption{The efficiency-corrected invariant mass distribution for the $J/\psi \omega$ final state. The solid line represents the total fit function. The dashed line is the background contribution. The solid histogram is the non $J/\psi \omega$ background estimated from sidebands. The vertical dashed line is placed at the X(3872) nominal mass.}
\end{figure}

\begin{figure}
  \includegraphics[height=.45\textheight]{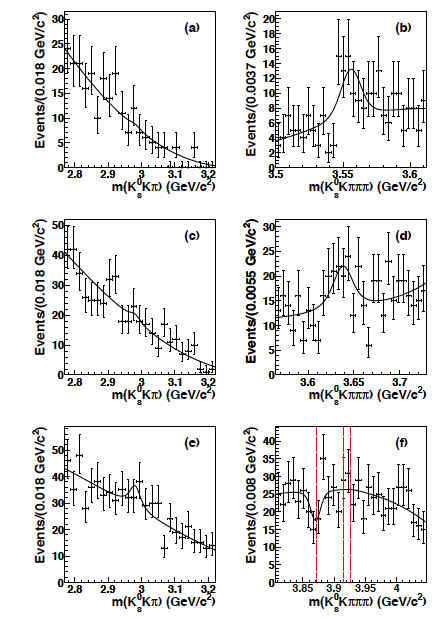}
  \label{fig2}
  \caption{Distribution of (a,c,e) $m(K^{0}_{s}K^{+}\pi^{-}\pi^{+}\pi^{-})$ with the fit function overlaid for the fit regions of the (a,b)$\chi_{c2}(1P)$, (c,d)$\eta_{c}(2S)$, and (e,f)~X(3872), X(3915) and $\chi_{c2}(2P)$. The vertical dashed line in (f) indicate the peak mass positions of the X(3872), X(3915), and $\chi_{c2}(2P)$.}
\end{figure}

\section{Study of the $\eta_{c}  \pi^{+}\pi^{-}$ in two-photon interactions.}
Studies of charmonium like-states in recent years have been performed using mainly the $J/\psi \pi^{+}\pi^{-}$ final state, no search using the 
 $\eta_{c}  \pi^{+}\pi^{-}$ final state has been conducted. Such a search may shed light on the quantum numbers of the charmonium-like-particle or the internal dynamics of these states. In particular, it has been suggested that if the  X(3872) is the $1^{1}D_{2}$ state $\eta_{c2}$ then the branching fraction ${\cal B}(X(3872) \to \eta_{c}\pi^{+}\pi^{-})$ could be significantly larger than ${\cal B}(X(3872) \to J/\psi \pi^{+}\pi^{-})$. The quantum numbers
 $J^{PC}=2^{-+}$ of the  $\eta_{c2}$ are consistent with the results of an angular analysis of $ X(3872) \to J/\psi \pi^{+}\pi^{-}$ \cite{Xang} and would allow production of X(3872) in two-photon fusion.
\babar   using 474 $\mathrm{fb^{-1}}$ studied the process $\gamma \gamma \to X \to \eta_{c}  \pi^{+}\pi^{-}$\cite{nir}
  where X stands for one of the resonance 
$\chi_{c2}(1P)$, $\eta_{c2}(2S)$, $X(3872)$,~$X(3915)$ or $\chi_{c2}(2P)$. The $\eta_{c}$ was reconstructed via its decay to $K_{S}^{0}K^{+}\pi^{-}$, with $K_{S}^{0} \to \pi^{+}\pi^{-}$. The signal yield for each X resonance is extracted from a two-dimensional fit to $m(K_{S}^{0}K^{+}\pi^{-})$ and $m(K_{S}K^{+}\pi^{-}\pi^{+}\pi^{-})$. Figure \ref{fig2} presents the two dimensional fits around each of the resonances. No significant signal is observed in any of the fits. Table \ref{fig3} summarizes all the fit result. Upper limits are obtained on the branching fraction ${\cal B}(\eta_{c}(2S) \to \eta_{c}\pi^{+}\pi^{-}) <7.4 \%  ~@90 \% {\mathrm C.L.})$ and ${\cal B}(\chi_{c2}(1P) \to \eta_{c}\pi^{+}\pi^{-}) <2.2 \% ~@90 \% {\mathrm C.L.})$

\begin{table}[htb]
\caption{Results of the $\gamma\gamma \to \eta_c\pi^+\pi^-$ fits. For each resonance $X$, we show the peak mass and 
width used in the fit and the 90\% CL upper limit on the product of the two-photon partial width $\Gamma_{\gamma\gamma}$ and the $X \to \eta_c\pi\pi$
branching fraction. For the $X(3872)$ and the $X(3915)$ we assume J = 2.}
\begin{tabular}{|c|c|c|c|c|} 
\hline
Resonances & $M_X$ (MeV/c$^2$) & $\Gamma_X$ (MeV) & UL $\Gamma_{\gamma\gamma}$ (eV) \\ 
\hline
$\chi_{c2}(1P)$  	& $3556.20\pm0.99$		& $1.97\pm0.11$	&15.7 \\
$\eta_c(2S)$		& $3638.5\pm1.7$		& $13.4\pm5.6$	&133 \\
$X(3872)$		& $3871.57\pm0.25$		& $3.0\pm2.1$	&11.1 \\
$X(3915)$		& $3915.0\pm3.6$		& $17.0\pm10.4$	&16 \\
$\chi_{c2}(2P)$	& $3927.2\pm2.6$		& $24\pm6$		&19 \\
\hline
\end{tabular}
 \label{fig3}
\end{table}

 \section{Search for the $Z_{1}(4050)^{+}$ and $Z_{2}(4250)^{+}$}
\begin{figure}
 \includegraphics[height=.3\textheight]{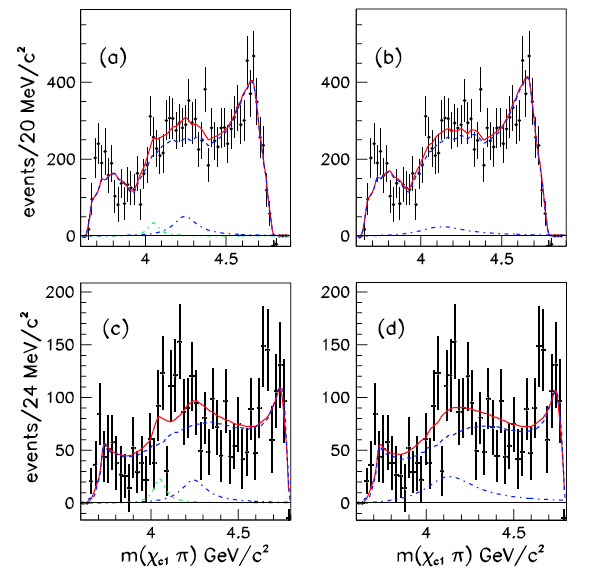}
  \label{fig4}
  \caption{(a),(b) Background-subtracted and efficiency-corrected $\chi_{c1}\pi$ mass distribution for $B\to \chi_{c1}K\pi$. (a) Fit with the $Z_{1}(4050)^{+}$ and $Z_{2}(4250)^{+}$ resonances. (b) Fit with only the $Z_{1}(4050)^{+}$ resonance. (c),(d) Efficiency-corrected and background-subtracted $\chi_{c1}\pi$ mass distribution in the $K\pi$ mass region where Belle found the maximum resonance activity: $1.0~<~m^{2}(K\pi)~<~1.75{\mathrm GeV^{2}/c^{4}}$. (c) Fit with $Z_{1}(4050)^{+}$ and $Z_{2}(4250)^{+}$ resonances. (d) Fit with only the $Z(4150)^{+}$ resonance. The dot-dashed curves indicate the fitted resonant contributions.}
  \end{figure}

In 2008 the Belle Collaboration reported the observation of a resonance-like structure called the $Z(4430)^{+}$ decaying to $\psi(2S)\pi^{+}$ in the study of the process $B\to \psi(2S)K\pi$ \cite{zbelle}. This claim generated a great deal of interest \cite{maiani},~\cite{karlip} since such states must have a minimum quark content $c\bar{c}\bar{d}u$, and thus would represent an unequivocal manifestation of four-quark meson state. The \babar collaboration searched the  $Z(4430)^{+}$ in a similar analysis of the Belle collaboration in the process  $B\to \psi(2S)K\pi$ and also in $B\to J/\psi K\pi$ \cite{zbabar}
but they did not find any structure neither in the $J/\psi \pi$ nor in the $\psi(2S) \pi$. Recently Belle performed an amplitude analysis of the $J/\psi K\pi$ \cite{zcharm} system finding no significant evidence of the $Z(4430)^{+}$ in agreement with \babar. In 2009 the Belle Collaboration reported also the observation of two resonance-like structures similar to the $Z(4430)^{+}$ in the study of the $\bar{B}^{0}\to \chi_{c1}K^{-}\pi^{+}$ \cite{z12belle}. These new structures were labeled as the $Z_{1}(4050)^{+}$ and $Z_{2}(4250)^{+}$, both decaying to $\chi_{c1}\pi^{+}$. \babar using a data sample of 429 $\mathrm{fb^{-1}}$ has recent searched for the $Z_{1}(4050)^{+}$ and $Z_{2}(4250)^{+}$in the process $\bar{B}^{0}\to \chi_{c1}K^{-}\pi^{+}$ and in the decay $B^{+}\to K_{s}^{0}\chi_{c1}\pi^{+}$ \cite{z12babar}; with the $\chi_{c1} \to J/\psi \gamma$. In the \babar analysis the $\chi_{c1}\pi^{+}$ mass distribution, after background subtraction and efficiency-correction, has been modeled using the angular information from the $K\pi$ mass distribution that has been represented using the Legendre polynomial moments. 
The excellent description of the $\chi_{c1}\pi^{+}$ mass distribution given by this analysis approach shows that there is no need for any additional 
resonance to model the distribution.  
Figure  \ref{fig4} shows the result of the fit done on the $\chi_{c1}\pi^{+}$ mass spectrum using two or one scalar Breit-Wigners with parameters fixed to the Belle measurement. In all the fit cases there are  no significant resonant structure, the statistical significance obtained is very low $<2 \sigma$. The ULs on the 90 \% CL on the branching fractions are for the one resonance fit:
${\cal B}(\bar{B}^{0} \to Z^{+}K^{-}) \times {\cal B}(Z^{+} \to \chi_{c1}\pi^{+})<4.7 \times 10^{-5}$  while for the two resonances fit are: 
${\cal B}(\bar{B}^{0} \to Z_{1}^{+}K^{-}) \times {\cal B}(Z^{+}_{1} \to \chi_{c1}\pi^{+})<1.8 \times 10^{-5}$ and ${\cal B}(\bar{B}^{0} \to Z^{+}_{2}K^{-}) \times {\cal B}(Z^{+}_{2} \to \chi_{c1}\pi^{+})<4.0 \times 10^{-5}$.

\section{Study of the $J/\psi\pi^{+}\pi^{-}$ system via Initial State Radiation (ISR).}
\begin{figure}
 \includegraphics[height=.2\textheight]{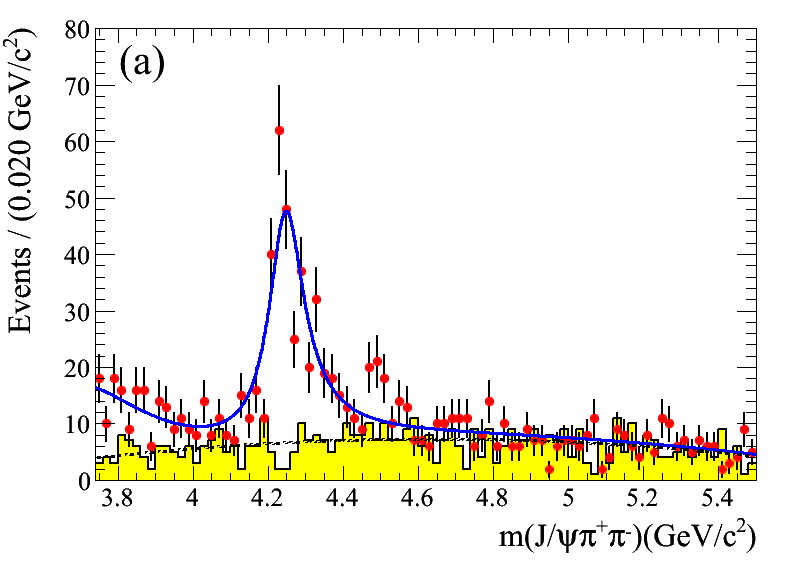}
 \includegraphics[height=.2\textheight]{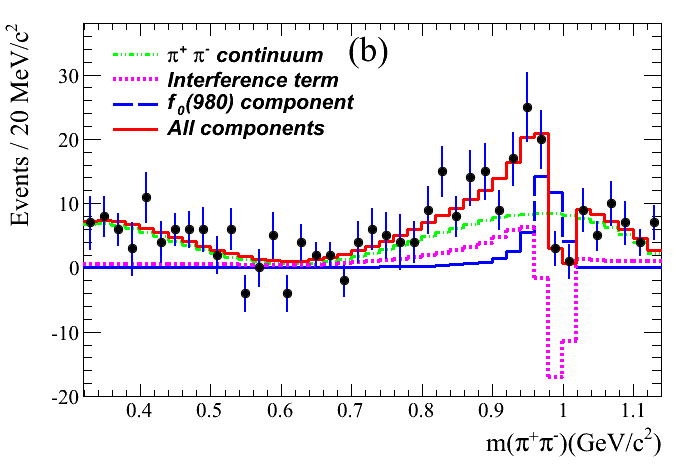}
  \label{fig5}
  \caption{(a): the $J/\psi \pi^{+}\pi^{-}$ mass spectrum from 3.74 ${\mathrm GeV/c^{2}}$ to 5.5 ${\mathrm GeV/c^{2}}$; the point represent the data and the shaded histogram is the background from the $J/\psi$~sideband; the solid curve represent the fit result. (b) the $\pi^{+}\pi^{-}$ distribution of the  from the Y(4260) decay  to $J/\psi \pi^{+}\pi^{-}$. The solid curve represent the fit using the model described in the text.}
    \end{figure}

In 2005 \babar discovered the Y(4260) in the process $e^{+}e^{-} \to \gamma_{ISR} Y(4260)$, with the $Y(4260) \to J/\psi\pi^{+}\pi^{-}$ \cite{Ybabar}.
Since it is produced directly in $e^{+}e^{-}$ annihilation it has $J^{PC}=1^{--}$.
The observation of the decay
mode $J/\psi\pi^0\pi^0$ \cite{Ypi0} established that it has zero isospin. 
However it is not observed to decay to $D^*\bar{D^*}$ \cite{Ydd}, nor to $D_s^*\bar{D^*_s}$ \cite{Yds}, so that its properties do not lend 
themselves to a simple charmonium interpretation, and its nature is still unclear.
A subsequent Belle analysis \cite{Ybelle} of the same final state suggested also the existence
of an additional resonance around 4.1 GeV/c$^2$ that they dubbed the Y(4008).
\babar has performed recently a new analysis \cite{Ybabarnew}
 of this process using 454 ${\mathrm fb^{-1}}$. In this new study the region below 4.0 ${\mathrm GeV/c^{2}}$ has been studied for the first time. As shown on Figure \ref{fig5}(a) in that region there is an excess of events above the $J/\psi$ sidebands background. To understand the nature of this contribution a detailed study of the $\psi(2S)$ line shape has been performed and the results is that it is not possible to discount the possibility that its due to  $J/\psi\pi^{+}\pi^{-}$  continuum cross section in this region. Figure \ref{fig5}(a) shows the fit to the $J/\psi\pi^{+}\pi^{-}$ mass distribution. A clear signal for the Y(4260) is seen; the values obtained from an unbinned-maximum-likelihood fit are: $m_{Y(4260)}=4244 \pm 5 \pm4~{\mathrm MeV/c^{2}}$, $\Gamma_{Y(4260)}=114^{+16}_{-15} \pm 7$ MeV and $\Gamma_{ee} \times {\cal B}(J/\psi \pi^{+}\pi^{-}) =9.2 \pm 0.8 \pm 0.7$eV. There is no evidence for the Y(4008) seen by Belle \cite{Ybelle}. \\ In the new \babar analysis a detailed study of the $\pi^{+}\pi^{-}$ system from the Y(4260) decay  to $J/\psi \pi^{+}\pi^{-}$ has been performed. The $\pi^{+}\pi^{-}$ mass distribution shown on Figure \ref{fig5}(b) seems to peak around the $f_{0}(980)$ mass; however the peak is displaced from the nominal $f_{0}(980)$ position, since it is around 940 ${\mathrm MeV/c^{2}}$. The fact that the peak is displaced and the particular shape of $m(\pi^{+}\pi^{-})$ distribution seems to suggest a possible interference between the  $f_{0}(980)$ and $m(\pi^{+}\pi^{-})$ continuum. To test this possibility the $f_{0}(980)$ line shape is taken from the \babar analysis of the $D_{S}^{+}\to \pi^{+}\pi^{-}\pi^{+}$ \cite{antimo} and this amplitude as been used in a very simple model to describe the $\pi^{+}\pi^{-}$ mass distribution: $|\sqrt{pol} +e^{i\phi}F_{f_{0}(980)}|^{2}$ where $pol$ is a polynomial function used to describe the  $m(\pi^{+}\pi^{-})$ continuum and 
$F_{f_{0}(980)}$ is the amplitude from the $D_{S}^{+}\to \pi^{+}\pi^{-}\pi^{+}$ \cite{antimo} analysis; $\phi$ allows for a phase difference between these amplitudes. The result of this study is shown on Figure \ref{fig5}(b) and indicates if there is a real $f_{0}(980)$ contribution to the decay of the Y(4260) to    $J/\psi\pi^{+}\pi^{-}$ its contribution is not dominant: $\frac{{\cal B}(Y_{4260} \to J/\psi f_{0}(980), ~f_{0}(980)\to \pi^{+}\pi^{-})}{{\cal B}(Y_{4260}   \to J/\psi \pi^{+}\pi^{-})}=(17 \pm 13) \%$.

  \begin{figure}
 \includegraphics[height=.18\textheight]{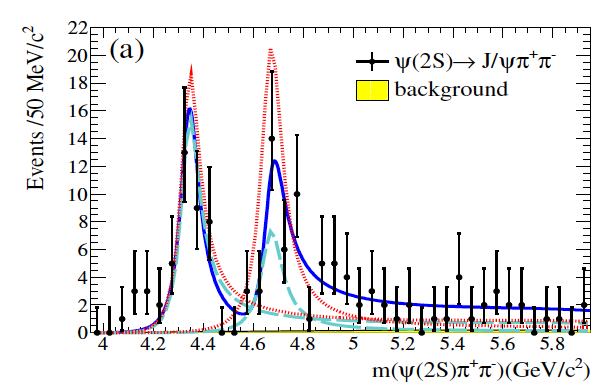}
  \label{fig6}
  \caption{The $\psi(2S)\pi^{+}\pi^{-}$ invariant mass distribution from threshold to 5.95 ${\mathrm GeV/c^{2}}$ for the $\psi(2S) \to J/\psi \pi^{+}\pi^{-}$; the points with error bars represent the data in the $\psi(2S)$ signal region, and the shaded histogram is the background estimated from the $\psi(2S)$ sideband regions. The solid curve show the result of the fit. }
  \end{figure}

\section{Study of the $\psi(2S)\pi^{+}\pi^{-}$ system via Initial State Radiation (ISR).}
In addition to the Y(4260), two more $J^{PC}=1^{--}$ states, the Y(4360) and the Y(4660) have been reported in ISR production $e^{+}e^{-}\to \psi(2S) \pi^{+}\pi^{-}$  \cite{Y2babar},~\cite{Y2belle}. While the  Y(4360) was discovered by \babar  \cite{Y2babar} and then confirmed by Belle ~\cite{Y2belle} the Y(4660) was only observed by the Belle Collaboration. \babar performed a new analysis using all its available dataset collected ad the $\Upsilon (nS)$,~n=2,3,4; that corresponds to an integrated luminosity of 520${\mathrm fb^{-1}}$. The $\psi(2S)\pi^{+}\pi^{-}$ mass spectrum for the $\psi(2S) \to J/\psi \pi^{+}\pi^{-}$ is reported in Figure~\ref{fig6} \babar observes two resonant structures, that have been interpreted as the Y(4360) and the Y(4660), respectively. The parameters values obtained from an unbinned-maximum-likelihood for the first resonance are $m_{Y(4360)}=4340 \pm 16 \pm 9~{\mathrm MeV/c^{2}}$, $\Gamma_{Y(4360)}=94 \pm 32  \pm 13$~MeV, and for the second one   $m_{Y(4660)}=4669 \pm 21 \pm 3~{\mathrm MeV/c^{2}}$, $\Gamma_{Y(4660)}=104 \pm 48 \pm 10$~MeV.





\bibliographystyle{aipproc}   




\end{document}